\documentclass[english]{article}
\usepackage[T1]{fontenc}
\usepackage[latin9]{inputenc}
\usepackage{amstext}
\usepackage{graphicx}
\usepackage{esint}

\makeatletter
\usepackage{authblk}

\makeatother

\usepackage{babel}
\begin{document}
\title{On the Impossibility of Supersized Machines}
\author[ ]{Ben Garfinkel}
\author[1]{Miles Brundage}
\author[2]{Daniel Filan}
\author[3]{Carrick Flynn}
\author[ ]{Jelena Luketina}
\author[ ]{Michael Page}
\author[3]{Anders Sandberg}
\author[3]{Andrew Snyder-Beattie}
\author[4]{Max Tegmark}
\affil[1]{School for the Future of Innovation in Society, Arizona State University}
\affil[2]{Department of Computer Science, University of California, Berkeley}
\affil[3]{Future of Humanity Institute, University of Oxford}
\affil[4]{Department of Physics, Massachusetts Institute of Technology}
\date{}
\setcounter{Maxaffil}{0}
\renewcommand\Affilfont{\itshape\small}

\date{April 1, 2017}
\maketitle
\begin{abstract}
In recent years, a number of prominent computer scientists, along
with academics in fields such as philosophy and physics, have lent
credence to the notion that machines may one day become as large as
humans. Many have further argued that machines could even come to
exceed human size by a significant margin. However, there are at least
seven distinct arguments that preclude this outcome. We show that
it is not only implausible that machines will ever exceed human size,
but in fact impossible. 
\end{abstract}

\section*{Introduction}

The history of life is often understood as a story of growth. If one
takes the long view, then one can trace an exponential curve from
our minuscule earliest ancestors, which were little more than self-replicating
molecules, to the substantial creatures that we are today (Payne,
2009). 

Although humanity became aware of this story only in the 19th century,
through the work of Charles Darwin, we have long had the privilege
of witnessing a partial recapitulation every time someone new comes
into the world (Darwin, 1859). Before each person is a full-sized
adult, they are first an invisibly small cell.

It is perhaps no surprise, then, that human largeness has for thousands
of years fascinated many of our greatest thinkers. While some have
sought to understand the nature and origins of largeness, others have
anxiously inquired: \textit{Could there ever be something larger than
a human?}

Evidence of this anxiety can be found as far back as humanity\textquoteright s
oldest recorded myth, \textit{The Epic of Gilgamesh}, in which the
monstrous giant Humbaba is appointed by Enlil, the king of the gods,
to terrorize mankind (Sandars, 1972). From this point onward, bellicose
giants have been a consistent presence in our literature, appearing
in works ranging from Homer\textquoteright s \textit{Odyssey} to the
English fairytale \textquotedblleft Jack and the Beanstalk'' (Homer,
1994; Anonymous, n.d.).

Over time, perhaps in response to our species\textquoteright{} growing
mastery of nature, it has become increasingly common to tell stories
in which people are the ones responsible for the larger-than-human
(or \textquotedblleft supersized\textquotedblright ) creatures that
threaten them. For generations, audiences have been drawn to tales
of frightful creations such as Frankenstein\textquoteright s monster,
described as over eight feet tall and \textquotedblleft proportionally
large'', and the golems of Kabbalah, which some rabbis feared would
grow large enough to destroy the universe (Shelly, 2008; Moshe, 1990).

This archetype has perhaps never been more prevalent than it is in
modern Hollywood films, however. Inspired by the apparently steady
march of technological progress, and the wild speculations of futurists,
our media has become saturated with images of murderous supersized
machines.

In the long-running \textit{Transformers} film series, machines known
as Decepticons, each perhaps the size of a hundred men, repeatedly
threaten to exterminate humanity with their enormous metal bodies
(Bay, 2007). Numerous entries in the \textit{Godzilla} film series
feature machines so large that they can crush portions of the Tokyo
skyline with a single step (Honda, 1975). We find that the \textit{Matrix}
film series, the \textit{Terminator} film series (notable for its
casting of an exceptionally large actor), and countless others also
feature supersized machines that seek to cause the extinction of the
human species (The Wachowskis, 1999; Cameron, 1984).

It does not help that in recent years a number of computer scientists,
philosophers, and other academics have publicly lent credence to the
possibility of supersized machines. There has been no shortage of
media coverage of these figures' pronouncements.\footnote{In addition, it has become very common for articles on recent trends
in computer science to use terms such as ``big data'' and ``massive
neural networks'' in ways that are likely to be misinterpreted. Reading
these articles, even ones that appear in highly reputable newspapers,
it is often unclear whether their authors are aware that the use of
size language in these contexts is purely metaphorical.}

However, perhaps fortunately, all predictions of a coming age of supersized
machines are fundamentally misguided. We present seven distinct arguments,
each of which suffices to show that supersized machines are impossible.\footnote{It is worth clarifying that there are, of course, systems today that
appear to exceed human size in narrow dimensions. Lamp posts are one
example. The predictions that we are considering concern some more
general notion of largeness.} 

\section*{Arguments Against Supersized Machines}

\subsubsection*{1. The Irreducible Complexity of the Human Body}

Despite having been an active research area for hundreds of years,
developmental biology has hardly progressed beyond its initial stages.
We are far from being able to tell a story in all but the bluntest
of terms of how a human zygote is able to transform itself, over the
course of two decades, into an adult that is several orders of magnitude
larger (Cameron, 2012).

Scientists are at the point of being able to identify traits that
correlate with largeness\textemdash certain genetic markers, for instance\textemdash but
they have nothing like a complete theory of the causal pathways that
explain these correlations. All attempts to construct such a theory
have been stymied by the irreducible complexity of the human body,
which contains tens of thousands of distinct proteins (Wilhem, 2014).
It seems inevitable that, for this same reason, all future attempts
will fail as well.

Since we cannot comprehend the processes responsible for human largeness,
it follows that we will never be able to produce machines that surpass
this largeness. 

\subsubsection*{2. The Meaninglessness of \textquotedblleft Human-Level Largeness\textquotedblright{}}

One simple reason that we can reject predictions of supersized machines
is that these predictions are not in fact well-formed. 

The term \textquotedblleft supersized machine\textquotedblright{}
implies a machine that has crossed some threshold, which is often
denoted \textquotedblleft human-level largeness.\textquotedblright{}
However, it is not clear what \textquotedblleft human-level largeness\textquotedblright{}
could refer to. Has a machine achieved human-level largeness if it
has the same height as the average human? If it has the same volume?
The same weight? Or some more complex trait, perhaps the logarithm
of girth multiplied by the square of height?\footnote{Note also that humans vary quite significantly along all of these
dimensions, and that even among humans there is no single accepted
measure of largeness (Pomeroy, 2015).}

When one begins to consider these questions, one quickly concludes
that there are an infinite number of metrics that could be used to
measure largeness, and that people who speak of \textquotedblleft supersized
machines\textquotedblright{} do not have a particular metric in mind.
Surely, then, any future machine will be larger than humans on some
metrics and smaller than humans on others, just as they are today.

One might say, to borrow Wolfgang Pauli\textquoteright s famous phrase,
that predictions of supersized machines are \textquotedblleft not
even wrong'' (Peierls, 1960).

\subsubsection*{3. The Universality of Human Largeness}

A further reason why it is senseless to speak of machines that are
larger than people is that humans already possess the property of
universal largeness.

By this, we mean that humans are capable of augmenting their bodies
or coming together to become indefinitely large, no matter the metric
chosen. If a human would like to be taller, they can stand on a chair
or climb onto another human\textquoteright s shoulders. If they would
like to be wider, they can begin consuming a high-calorie diet or
simply put on a thick sweater (Hensrud, 2004; Figure 1). There are
recorded cases of humans joining their bodies together to reach heights
of up to 12 meters (Guinness World Records, 2013).
\begin{figure}[t]
\begin{centering}
\includegraphics[width=0.38\paperwidth]{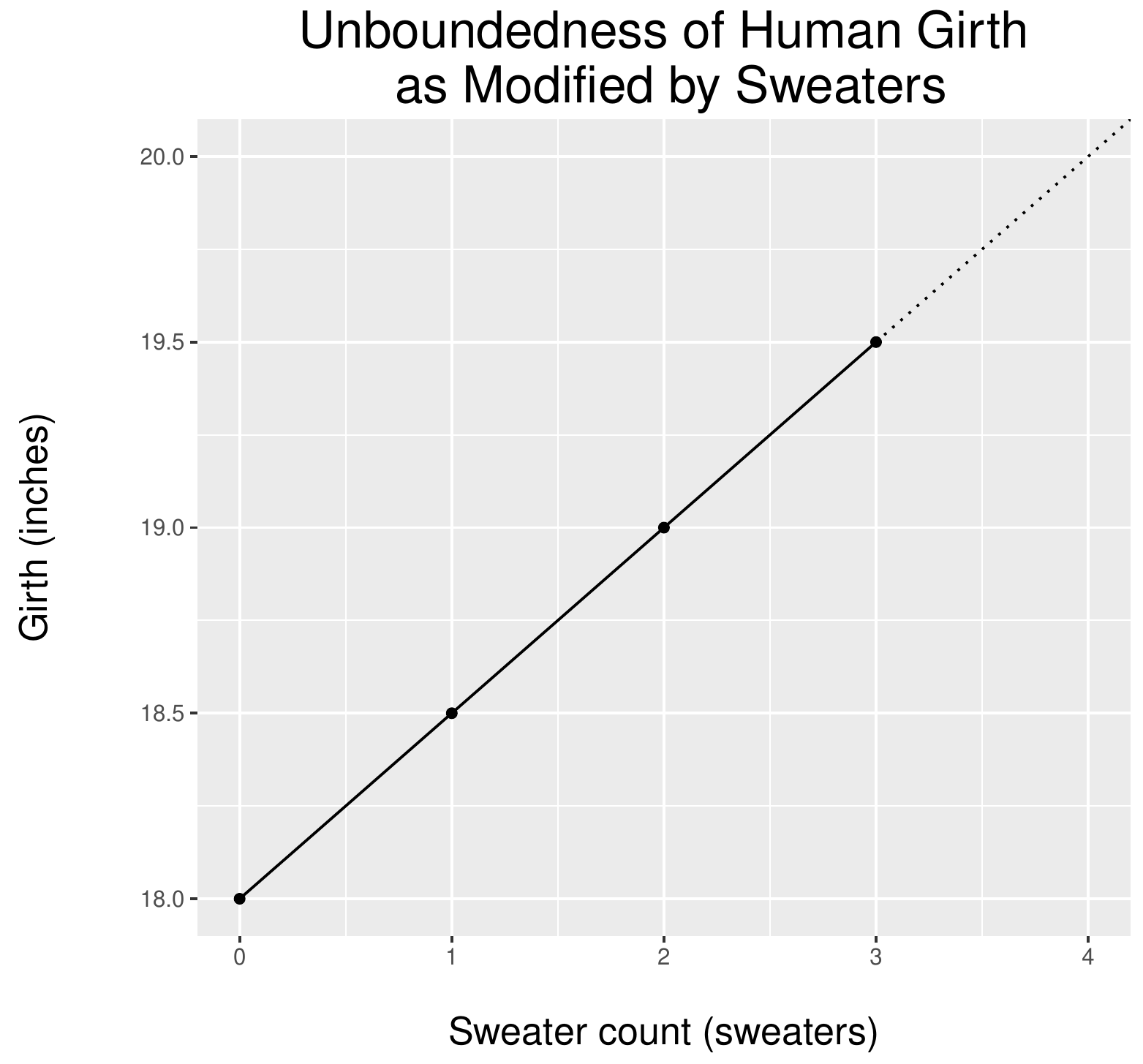}
\par\end{centering}
\caption{There is no upper bound on how large humans can be (Gonzalez, 2017).}
\end{figure}

In short, since there exists no physical law to put an upper bound
on human largeness, humans can be of any size. It follows, then, that
no machine could ever really be larger than a human. 

\subsubsection*{4. The Psychological Origins of Belief in Supersized Machines}

By explaining why some people may be inclined to worry about supersized
machines, evolutionary psychology reveals that such fears are not
rational.

It is only natural that our ancestors should have developed a fear
of beings larger than themselves. The greater a tribe member\textquoteright s
size, the more capable they are of employing violent coercion against
other members or stealing their mates (Brewer, 2009). For this reason,
vigilance toward the possibility of very large things was a highly
advantageous trait.

Although largeness now plays a much-diminished role, at least in Western
societies, there has been little time for human psychology to adapt
(Donald, 1993). Furthermore, given the central role that technology
plays in modern life, we should find it perfectly unsurprising that
many people (especially ``alpha males'' enmeshed in Silicon Valley
culture) have come to possess a fear of supersized machines.

Thus it is evolution, rather than logic or evidence, that serves as
the true source of the belief that supersized machines are possible.
It follows that we can safely assume this belief to be false.

\subsubsection*{5. Humans and Machines Together Will Always Be Larger Than Machines
Alone }

When writers discuss the possibility of supersized machines, they
appear to be missing a crucial consideration: No machine could ever
be larger than that same machine and a human together.

If machines are to play a role in pushing forward the frontier of
largeness, then this role could only ever be to supplement human largeness. 

This is another simple reason why it is senseless to imagine larger-than-human
machines.\footnote{This consideration also suggests that credible machine largeness researchers
ought to focus on human-machine interfaces, which enable size-enhancing
machines to be attached directly to the human body. Existing work
on stilts may suggest one promising research direction (Smith, 2010). } 

\subsubsection*{6. The Hard Problem of Largeness}

Suppose one were to concede that machines could become as large as
humans, in some sense related to physical extension (although this
is of course impossible).

Even if this were so, there would still remain a second, more meaningful
sense of the word \textquotedblleft large\textquotedblright{} that
would not apply to these machines.

This second kind of largeness is the one evoked whenever someone is
described as \textquotedblleft larger than life\textquotedblright{}
or \textquotedblleft living large'' (Tom, 2004). Largeness of this
sort is a non-physical (i.e. non-natural) property, separate from
the mundane physical property that \textquotedblleft largeness\textquotedblright{}
most often denotes. 

To build a large machine, then, in the meaningful sense, we would
first need to solve the \textquotedblleft hard problem\textquotedblright{}
of determining what this non-physical property is and how it arises.
However, it is not at all clear that the problem is soluble, since
the traditional methods of science seem equipped only to deal with
questions that concern the physical world (Hall, 2010). Furthermore,
the notion of a machine ``living large'' strikes one as intuitively
implausible (perhaps even absurd).

Therefore, machines will never \textit{truly} be large. 

\subsubsection*{7. Quantum Mechanics and Gödel\textquoteright s First Incompleteness
Theorem}

Quantum theory, as traditionally formulated, divides the world up
into microsystems and macrosystems (Heisenberg, 1949). Within microsystems
lie small objects, such as particles, and within macrosystems lie
large objects, such as humans. 

The theory tells us that objects in microsystems may initially have
no definite properties at all, such that any question concerning a
given particle\textquoteright s position, momentum, and so forth,
will simply lack an answer. However, the remarkable ability that humans
possess, as a result of their largeness, is the ability to force objects
in microsystems to take on definite properties by performing \textquotedblleft measurements\textquotedblright{}
on them. For example, if a human \textquotedblleft measures\textquotedblright{}
that a particle has a certain location, then it becomes a new fact
that the particle has this location.

One of the great mysteries of quantum mechanics, that its originators
never succeeded in resolving, is the question of what distinguishes
microsystems from macrosystems (Bell, 1990). It seems that we are
to understand that some fundamental line separates the large from
the small, such that small objects exist in a sort of limbo until
large objects perform measurements on them. However, we lack guidance
on how to draw this line, and it is difficult to understand how and
why the line exists at all. The problem of making sense of this line,
and thereby uncovering the nature of largeness, is known as \textquotedblleft the
measurement problem.\textquotedblright{}

A partial answer to the measurement problem may be suggested by Kurt
Gödel\textquoteright s first incompleteness theorem (Gödel, 1931).
This theorem was first proved in 1931, although its full significance
arguably remains to be appreciated. The theorem states that, for any
sufficiently expressive formal system, the system must either be inconsistent
or incapable of proving true or false all statements that are expressible
within the system.

To understand how Gödel\textquoteright s theorem can resolve the measurement
problem, it is perhaps most useful to apply the lens of quantum stochastic
calculus (Kholevo, 1991). QSC, as a reminder, generalizes classical
stochastic calculus to cover cases of non-commuting random variables,
which are ubiquitous in quantum mechanics. Take the quantum Stratonovich
integral of a system operator, \textit{g(t)}, which is given by (Gardener,
2004):

\[
(S)\intop_{t_{0}}^{t}g(t')dB(t')=lim_{n\rightarrow\text{\ensuremath{\infty}}}\sum_{i=1}^{n}\frac{g(t_{i})+g(t_{i+1})}{2}(B(t_{i+1},t_{0})-B(t_{i},t_{0}))
\]

Applying this expression, it is trivial to show that:

\[
(S)\intop_{t_{0}}^{t}g(t')dB(t')-(S)\intop_{t_{0}}^{t}dB(t')g(t')=\frac{\sqrt{\text{\ensuremath{\gamma}}}}{2}\intop_{t_{0}}^{t}dt'[g(t'),c(t')]
\]

Now suppose that we would like to formalize this deduction within
non-well-founded set theory, to which Gödel's theorem of course applies
(Aczel, 1988). Importantly, by assuming the axiom of anti-foundation
we are able to introduce into our analysis self-referential objects,
such as Quine atoms, which possess the property of being large enough
to contain themselves.

Although the technical details from this point onward are unfortunately
too dense to include in a general-audience essay of this sort, assuming
as they do familiarity with constructive non-standard analysis, it
suffices to say that supersized machines cannot be made (Figure 2).
\begin{figure}[t]
\begin{centering}
\includegraphics[width=0.38\paperwidth]{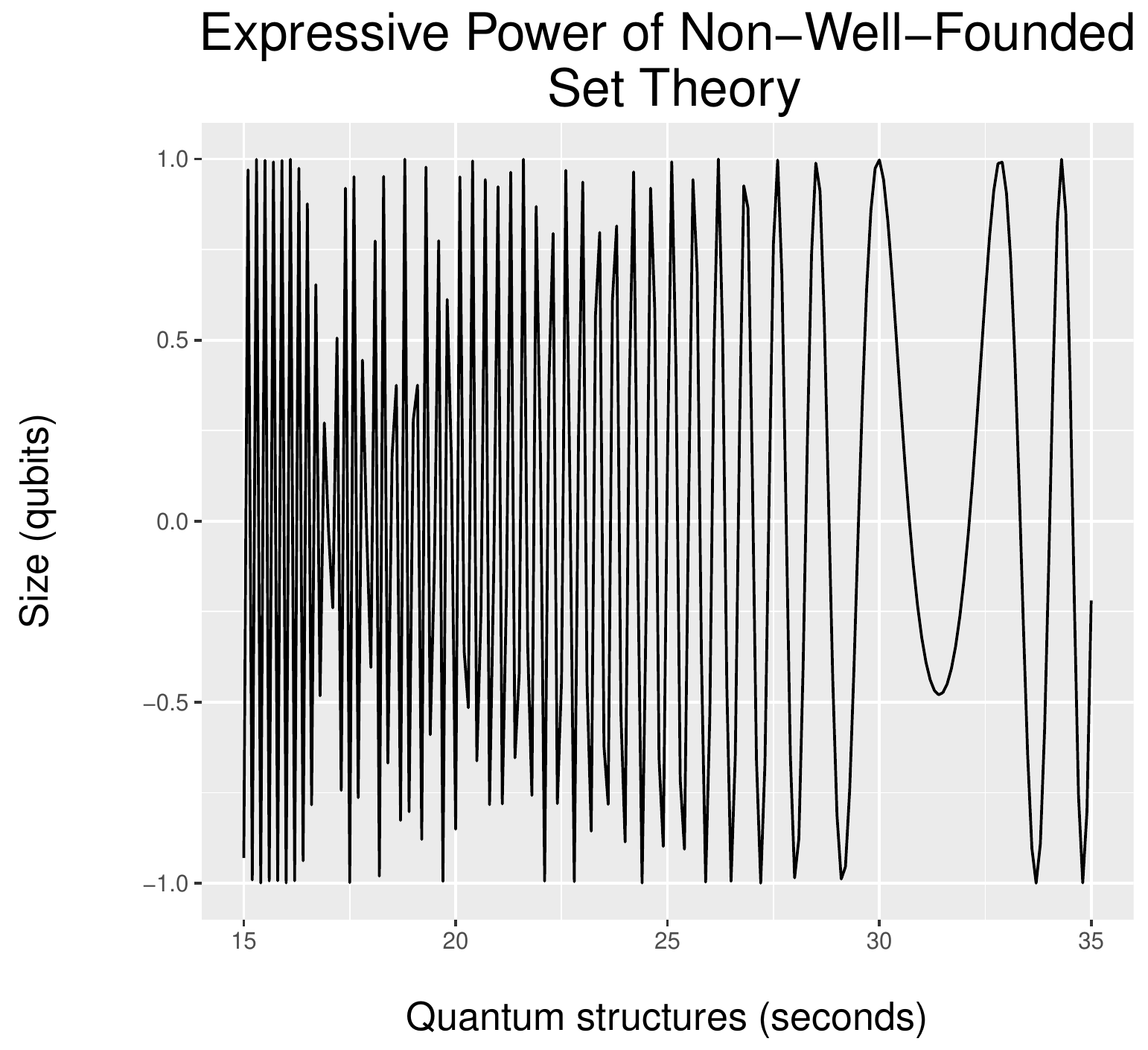}\caption{Machines cannot be large (Gonzalez, 2017).}
\par\end{centering}
\end{figure}

\section*{Conclusion}

We have presented seven distinct arguments against the possibility
of supersized machines. While each of these arguments would be sufficient
on its own, the conjunction of them surely constitutes an insurmountable
barrier to the belief that an age of supersized machines lies anywhere
on the horizon.\footnote{One may wonder why we have felt it necessary to demonstrate that supersized
machines are impossible, rather than arguing for the much weaker claim
that supersized machines are unlikely to be developed soon. The reason
is that, counter-intuitively, many of the academics who have expressed
concern about supersized machines appear to accept this weaker claim.
They argue from the position, currently controversial among policy-makers,
that it is worth preparing for distant or low-probability events (Bedford,
2001). This position has led many to stake out similarly provocative
stances in favor of climate change mitigation, pandemic preparedness,
and seatbelt use.}

Our conclusion is in at least one way a relief. There is no reason
to fear preposterous stories about towering Terminator machines.

However, our conclusion might also be taken as a sad one. We are the
largest things in the universe, and we will never be otherwise. 

Fantasies of supersized machines hold an appeal, in addition to inspiring
fear, because it is tempting to imagine these machines as perfected
versions of ourselves. They are who people would be if only we were
a little larger. They are steadier, and more able to look down upon
the world with a distant wisdom, rather than becoming entangled in
the insignificant details close to ground.

It can be nice to think that if we are unable to resolve our own problems
here on Earth, then maybe this is only because we lack the size.

A world in which we are the largest things conceivable is a world
without excuses. We submit that this is a good thing, however. It
is time to stop daydreaming about something larger than ourselves,
and time to begin understanding how large we truly are.

\end{document}